\documentclass[aps,prl,reprint,showpacs,twocolumn,superscriptaddress]{revtex4}

\usepackage[dvips]{graphicx}
\usepackage{dcolumn}% Align table columns on decimal point
\usepackage{bm}% bold mat/cite
\usepackage{xspace}
\usepackage{multirow}
\usepackage{natbib}
\usepackage{color}

\begin{document}

\preprint{}
%\title{Room Temperature Coherence and Full Spin-Polarization of Topological Surface States on Bi$_2$Se$_3$}
\title{Persistent Coherence and Spin-Polarization of Topological Surface States on Topological Insulators}
\author{Z.-H. Pan}
\affiliation{Condensed Matter Physics and Materials Science Department, Brookhaven National Lab, Upton, NY 11973, USA}
\author{E. Vescovo}
\affiliation{National Synchrotron Light Source, Brookhaven National Lab, Upton, NY 11973, USA}
\author{A. V. Fedorov}
\affiliation{Advanced Light Source, Lawrence Berkeley National Laboratory, Berkeley, CA 94720, USA}
\author{G. D. Gu}
\author{T. Valla}
\email{valla@bnl.gov}
\affiliation{Condensed Matter Physics and Materials Science Department, Brookhaven National Lab, Upton, NY 11973, USA}
\date{\today}

\begin{abstract}
Gapless surface states on topological insulators are protected from elastic scattering on non-magnetic impurities which makes them promising candidates for low-power electronic applications. However, for wide-spread applications, these states should remain coherent and significantly spin polarized at ambient temperatures. Here, we studied the coherence and spin-structure of the topological states on the surface of a model topological insulator, Bi$_2$Se$_3$, at elevated temperatures in spin and angle-resolved photoemission spectroscopy. We found an extremely weak broadening and essentially no decay of spin  polarization of the topological surface state up to room temperature. Our results demonstrate that the topological states on surfaces of topological insulators could serve as a basis for room temperature electronic devices.

\end{abstract}
\vspace{1.0cm}


\maketitle
\pagebreak
Three-dimensional topological insulators (TIs) have Dirac-like surface states in 
which the spin of the electron is locked perpendicular to its momentum
 in a chiral spin-structure where electrons with opposite momenta
have opposite spins \cite{Fu2007a,Noh2008a,Hsieh2008,Zhang2009,Hsieh2009,Xia2009,Chen2009,Pan2011c}.
A direct consequence of this spin-structure is that a 
backscattering is not allowed if a time-reversal-invariant perturbation, such as non-magnetic disorder, is present \cite{Fu2007a}, making these topological surface states (TSSs) extremely robust and promising candidates for spintronics and quantum computing applications. \cite{Biswas2010,Fu2009,Guo2010,Liu2009,Zhou2009}. The measured degree of the state\rq{}s spin polarization was very low in the early spin- and angle-resolved photoemission spectroscopy (SARPES) experiments $P\sim0.2$ \cite{Hsieh2009}, but more recent measurements show much larger, almost full polarization at low temperatures \cite{Pan2011c,Jozwiak2011}. These new results are in stark contrast with theoretical models that predict a significant reduction of the intrinsic polarization due to the strong spin-orbit entanglement \cite{Yazyev2010} and propose that the high degree of measured polarization is an artifact due to the polarization of light used in SARPES experiments \cite{Park2012a}. In addition, recent \lq\lq{}layer-resolved\rq\rq{} calculations have suggested that the measured polarization has to be reduced due to the reversal in the topmost atomic layer \cite{Henk2012,Eremeev2012}. If correct, these models would place serious limitations on the  applicability of topological insulators in electronic and spintronic devices.
%fig 1 %#######################################################################
\begin{figure*}[htb]
\begin{center}
\includegraphics[width=12cm]{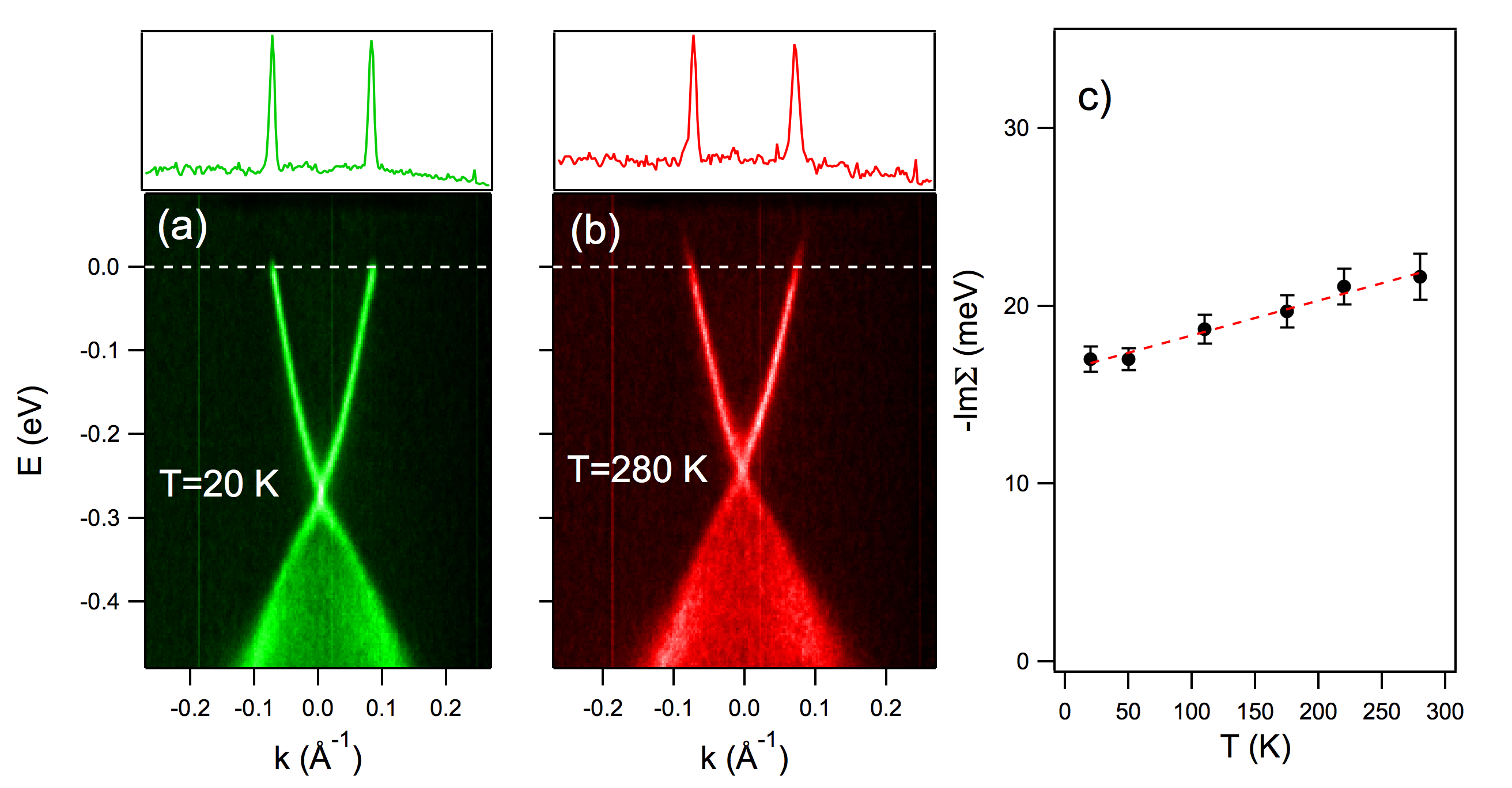}
\caption{Temperature effects on the spectra of topological surface state on Bi$_2$Se$_3$. (a) ARPES intensity along the $\Gamma$K line in the surface Brillouin zone at 20 K and (b) at 280 K. The top panels show the MDCs at the Fermi level, with two Lorentzian shaped peaks, corresponding to $\pm k_F$. 
 (c) Temperature dependence of Im$\Sigma(0)$ for TSS on Bi$_2$Se$_3$. Im$\Sigma(0)$ is obtained  by fitting the MDC peaks with Lorentzian line-shape. The error bars represent standard deviation from the fitting. Dashed line is the linear fit of Im$\Sigma(0)$.
}
\label{Fig1}
\end{center}
\end{figure*}
%#######################################################################
The second aspect, important for applications is the states\rq{}s capacity to remain coherent over long distances and resist the scattering. The elastic scattering has been explored in scanning tunneling microscopy (STM) experiments \cite{Roushan2009,ZhangSTM2009,Alpichshev2010,Seo2010,Hanaguri2010} where it has been shown that backscattering is indeed strongly suppressed or completely absent, despite strong atomic scale disorder. 
Angle-resolved photoemission spectroscopy (ARPES) studies have indicated that the topological state is remarkably insensitive to both non-magnetic and magnetic impurities in the low doping regime, where the Fermi surface (FS) is nearly circular \cite{Valla2012a}. Question of inelastic scattering and in particular the strength of electron-phonon coupling is also very important as electron mobilities, important for functioning of electronic devices, are limited at elevated temperatures by scattering on phonons. According to the recent theoretical studies \cite{Giraud2011}, topological states on TIs should be relatively strongly coupled to phonons. The experimental situation is somewhat inconclusive with the He-scattering experiments \cite{Zhu2011a} indicating relatively strong coupling, while the ARPES studies report very inconsistent results for the coupling strength, ranging from extremely weak ($\lambda\sim 0.08$) \cite{Pan2012,Park2011}, to extremely strong ($\lambda\sim 3$) \cite{Kondo2013} coupling in the same material. 
Here, we show that the topological surface state retains its spin texture and remains fully coherent even at room temperature. This opens the prospect that topological insulators could serve as a basis for room temperature electronic devices.

The single crystals of Bi$_2$Se$_3$ were grown from high-purity (99.9999\%) elements by a modified floating zone method, where the Se-rich material was used in the melting
zone. The crystal growth rate was controlled at 0.5 mm per hour. The samples for ARPES and SARPES studies were cut from the same bulk piece and cleaved and measured in ultrahigh vacuum conditions (base pressure better than $2\times 10^{-9}$ Pa in both the ARPES and SARPES chambers). 
The ARPES experiments were carried out on a Scienta SES-100 electron spectrometer 
at the beamline 12.0.1 of the Advanced Light Source. The spectra were recorded at the photon energy of 50 eV, with the combined instrumental energy resolution of $\sim12$ meV and the angular resolution better than $\pm 0.07^{\circ}$. Samples were cleaved at low temperature (15-20 K) under ultra-high vacuum (UHV) conditions ($2\times10^{-9}$ Pa). The temperature was measured using a silicon sensor mounted near the sample.
The SARPES experiments were performed at the beamline U5UA at the National Synchrotron Light Source using the Omicron EA125 electron analyzer coupled to a mini-Mott spin polarimeter that enables measurements of the in-plane spin polarization. The spin-resolved data were recorded at room temperature, at 50 eV photon energy. Energy and angle resolution were approximately 40 meV and $0.5^{\circ}$, respectively. 

%fig 2 
%#######################################################################
\begin{figure*}[htb]
\begin{center}
\includegraphics[width=17cm]{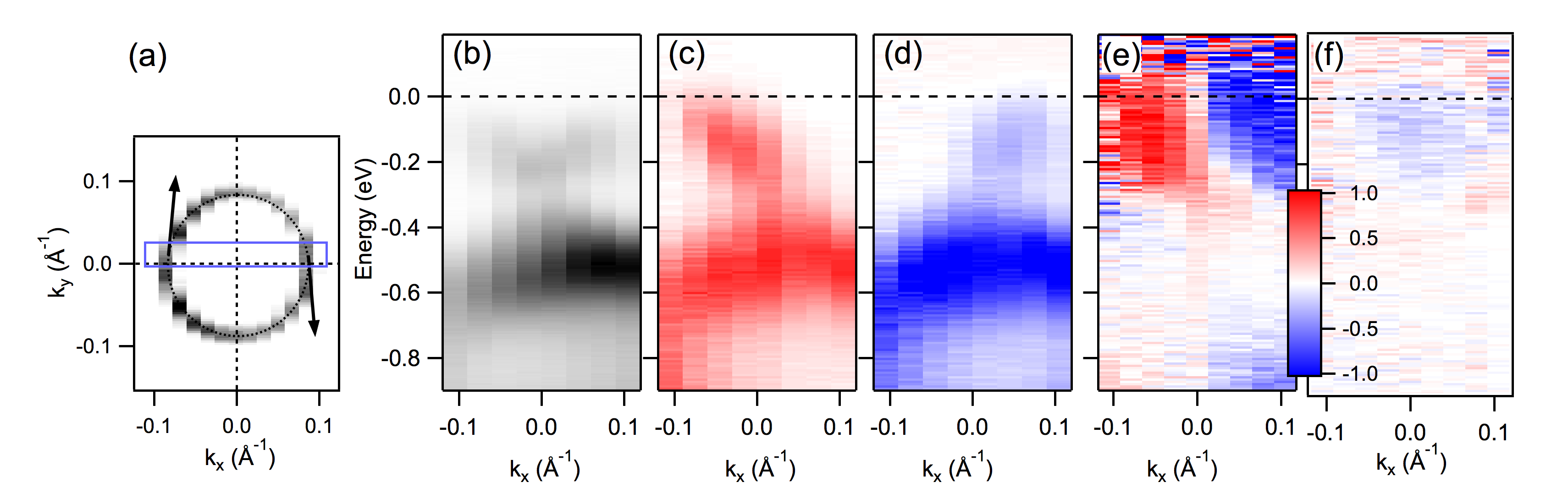}
\caption{Spin-Resolved ARPES spectra from Bi$_2$Se$_3$.  a) Fermi surface and b) spin-integrated photoemission intensity from the momentum region marked with the blue rectangle in a). c) Spin-Up and d) Spin-Down spectra from the same region, with the polarization vector along the $y$ direction. e) spin-polarization in the $y$ direction. f) spin-polarization in the $x$ direction. All the spectra were taken at 300 K with 50 eV photons. The arrows in a) represent the obtained spin direction.
}
\label{Fig2}
\end{center}
\end{figure*}
%#######################################################################
Fig. \ref{Fig1} illustrates the effects of raising temperature on the electronic structure of Bi$_2$Se$_3$ measured in ARPES around the center of the surface Brillouin zone. The rapidly dispersing conical state represents the TSS that forms a circular Fermi surface (Fig.3(a)). Its filling varies slightly with temperature as evident from the shift of the Dirac point from  $\approx0.27$ eV below the Fermi level at 20 K to $\approx0.23$ eV at 280 K \cite{Pan2012}. Upon raising the temperature, the TSS also seems to become marginally broader. The temperature induced shift and a slight change in the state\rq{}s width are fully reversible upon temperature cycling. Therefore, these effects reflect the intrinsic temperature induced changes in the quasi-particle dynamics.  

To quantify the changes in the spectral width of TSS, we have analyzed the photoemission spectra at different temperatures using the standard method where the momentum distribution curves (MDCs) are fitted with Lorentzian peaks \cite{Valla1999,Valla1999a}. The width of the Lorentzian peak, $\Delta k(\omega)$, is related to the quasiparticle scattering rate $\Gamma(\omega)=2|$Im$\Sigma(\omega)|=\Delta k(\omega)v_0(\omega)$, where $v_0(\omega)$ is the bare group velocity and Im$\Sigma(\omega)$ is the imaginary part of the complex self-energy. The most important observation here is that at the Fermi level, quasi-particle peaks show very little change between 20 K and 280 K (top panels in Fig. 1a), b)). Temperature broadening of a quasi-particle peak usually reflects an increase in the scattering on phonons and its near absence here points to a very weak coupling of TSS to phonons in Bi$_2$Se$_3$.
The electron-phonon coupling constant, $\lambda$, can be determined from the temperature slope of Im$\Sigma(0)$: Im$\Sigma(0,T)\approx \lambda\pi k_BT$, where $k_B$ is Boltzmann's constant. In panel c), we plot Im$\Sigma$ at $\omega=0$ as a function of temperature. Im$\Sigma$ increases very little with temperature. The linear fit gives $\lambda=0.072\pm 0.007$, amongst the weakest coupling constants ever reported \cite{Pan2012,Park2011}. Here, we want to emphasize that if the recent report \cite{Kondo2013} of an \lq\lq{}exceptionally large\rq\rq{} coupling, $\lambda\sim 3$, were correct, the TSS would be completely incoherent at room temperature, with $\Gamma$ approaching $\sim 0.5$ eV and the momentum width $\Delta k\sim 0.1$ \AA$^{-1}$, larger than $k_F$. Our Fig. 1(b) shows that this is clearly not the case.

%#######################################################################
\begin{figure}[htb]
\begin{center}
\includegraphics[width=8cm]{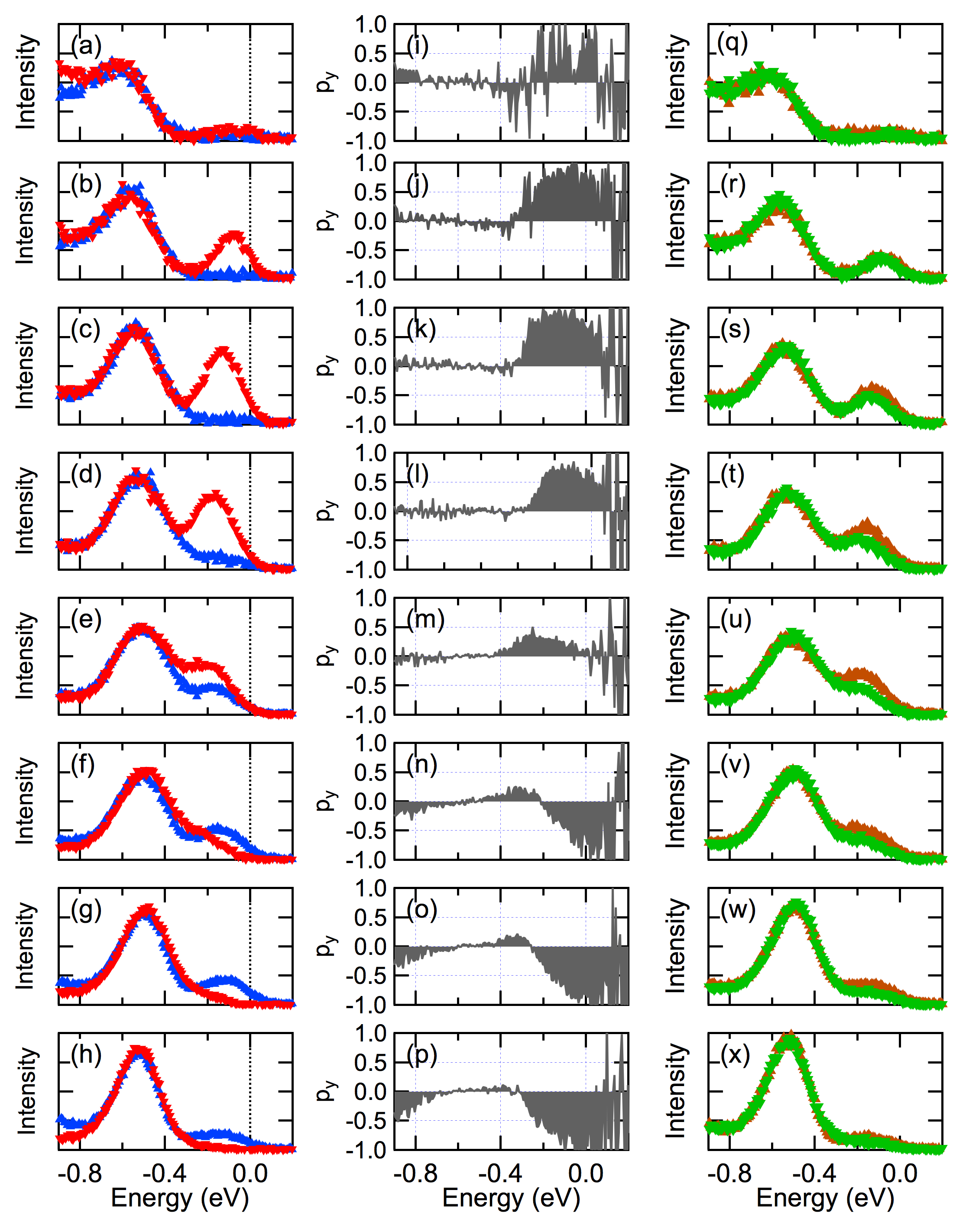}
\caption{Spin-Resolved ARPES spectra from Bi$_2$Se$_3$. a)-h) Spin-up (red) and spin-down (blue) EDCs and i)-p) spin-polarization along the $y$ direction, corresponding to the spectra from Fig. 3. q)-x) Spin-up (green) and spin-down (red) EDCs for the spin-polarization in the $x$ direction.
}
\label{Fig3}
\end{center}
\end{figure}
%#######################################################################
We note that the determining factor for surface transport is the surface state mobility, which can be expressed as $\mu_S=e\ell_{tr}/(\hbar k_F)$ for the Dirac-like carriers, where $\ell_{tr}$ represents the transport mean free path. $k_F$ and the quasiparticle mean-free path $\ell=(\Delta k)^{-1}$ are directly measured in ARPES from the position and the width of a Lorentzian peak in MDCs. As the transport in general is not sensitive to the small angle scattering events that may dominate quasiparticle scattering, $\ell_{tr}$ could be significantly longer than $\ell$, especially in systems in which the backscattering is strongly suppressed, as in the case of TSSs. Our results show that the mean free path does not decay with increasing temperature, implying that mobilities should also remain preserved. Therefore, the unperturbed and strongly coherent TSSs, as these studied here, have a strong potential to serve as a basis for room temperature electronic devices.

Another aspect of TSS, important for the high temperature (spin) electronic applications, is the degree of spin-polarization at elevated temperatures. In principle, an increase in various channels of inelastic scattering with temperature could depolarize the TSS and render it inadequate for large scale applications. For example, in ferromagnetic materials, the polarization decreases with temperature and vanishes completely above Curie temperature ($T_C$) due to the collapse of the exchange gap and/or due to the depolarization induced by the spin-flip scattering processes. Similarly, one  might also expect some temperature changes in the bulk electronic structure that could affect some of the parameters of the inversion gap and, consequently, the TSS. In contrast, we have found that TSS retains its spin texture and remains fully spin polarized at room temperature.
This is illustrated in Fig. 2, where we show the spin-resolved ARPES spectra of Bi$_2$Se$_3$ surface. The spin-integrated measurement of the Fermi surface (FS) (panel (a)) shows almost a perfectly circular contour of the FS. The spin-resolved measurements were obtained from the momentum line in the SBZ that was parallel to, but slightly displaced from the $K-\Gamma-K$ line, in steps of $0.5^{\circ}$. In panel (b) we show the spin integrated data along that momentum line, recorded with the spin-resolved analyzer. The spin-resolved spectra for the spin polarization in the $y$ direction are shown in panels (c) (spin up) and (d) (spin down). Panels (e) and (f) show polarizations in the $y$ and $x$ directions, respectively. It is obvious that the surface state has a full spin-up character in the $y$ direction for $k_x<0$ and spin-down character for $k_x>0$. Therefore, $P_y$ switches from +1 to -1 as the state disperses form $k_x<0$ to $k_x>0$ (Fig. 2(e)). We note that this is likely the highest polarization ever measured at room temperature in any system.

The same conclusion can be drawn from spin-resolved energy distribution curves, shown in Fig. 3, where it can be seen that the peak corresponding to the TSS is fully spin-polarized, with the spin in the $y$ direction. While there is a very prominent peak in the EDC for one spin direction, there is absolutely no intensity in the EDC for the opposite spin (panels (b)-(c) and (g)-(h) for example)! We note that due to a small but finite $k_y$, we also see a small polarization in the $x$ direction ($P_x\not= 0$), that does not switch the sign along the probed line (Fig. 2(f)). The observed $P_y$ and $P_x$ are consistent with the left-handed spin helicity of the TSS\rq{} Fermi surface.
These results show that the bulk gap remains large and inverted at the ambient temperature.  
In summary, we have shown that the TSS on the surface of Bi$_2$Se$_3$ stays coherent and that its spin texture remains intact up to ambient temperatures. This keeps the possibility that TSSs could serve as a basis for room-temperature devices open.

The work at Brookhaven is supported by the US Department of Energy (DOE) under Contract No. DE-AC02-98CH10886. 
ALS is operated by the US DOE under Contract No. DE-AC03-76SF00098.

%%%###############################
%\bibliography{TI_n}
\end{document}